\pdfoutput=1

\documentclass{article}
\usepackage[T1]{fontenc}
\usepackage[utf8]{inputenc}
\usepackage{ismir} 
\usepackage{amsmath,cite,url}
\usepackage{graphicx}
\usepackage{color}
\usepackage{multirow}
\usepackage{booktabs}
\usepackage{array}

\title{The SMC Blind Spot: A Failure Mode Analysis of State-of-the-Art Beat Tracking}

\threeauthors
  {Jaehoon Ahn} {Sogang University \\ \texttt{jahn@sogang.ac.kr}}
  {Tae Gum Hwang} {Sogang University \\ \texttt{academygum@sogang.ac.kr}}
  {Moon-Ryul Jung} {Sogang University \\ \texttt{moon@sogang.ac.kr}}

\def\authorname{J. Ahn, T. Hwang, and M. Jung}

\usepackage[bookmarks=false,pdfauthor={\authorname},pdfsubject={Beat Tracking Analysis},hidelinks]{hyperref}

\begin{document}

\maketitle

\begin{abstract}
Over the past two decades, the task of musical beat tracking has transitioned from heuristic onset detection algorithms to highly capable deep neural networks (DNN). Although DNN-based beat tracking models achieve near-perfect performance on mainstream, percussive datasets, the SMC dataset has stubbornly yielded low F-measure scores. By testing how well state-of-the-art models detect beats on individual tracks in the SMC dataset, we identify three distinct failure modes: octave errors, continuity errors, and complete tracking failure where all metrics fall below 0.3. We reveal that state-of-the-art models tend to generate "confident-but-wrong" activations. Furthermore, we show that the standard DBN's default minimum tempo of 55 BPM prevents it from inferring the correct tempo for 21\% of SMC tracks, forcing double-tempo predictions on slow music. By exposing such fundamental oversights, we provide concrete directions for improving beat and downbeat detection, specifically emphasizing training data diversification and multi-hypothesis tempo estimation.
\end{abstract}

\section{Introduction}\label{sec:introduction}

Beat tracking is a longstanding challenge in Music Information Retrieval (MIR) that involves estimating the temporal locations of musical beats in an audio signal. Modern approaches have progressed from onset detection functions \cite{scheirer1998tempo, dixon2001automatic, ellis2007beat, davies2007context, dixon2007evaluation, mcfee2014better} and recurrent neural networks \cite{bock2011enhanced, bock2014multi, bock2016joint, krebs2016downbeat} to temporal convolutional networks \cite{matthewdavies2019temporal, bock2020deconstruct, steinmetz2021wavebeat} and Transformer-based architectures \cite{hung2022modeling, zhao2022beat, cheng2023transformer, kim2023all, chang2024beast, foscarin2024beat}, with each generation improving scores on standard benchmarks. Nearly all current systems share a common pipeline: a neural network produces beat activation functions, which are then processed by a Dynamic Bayesian Network (DBN) \cite{whiteley2006bayesian, krebs2015efficient, bock2016joint, bock2016madmom} into a final beat sequence. On most evaluation datasets, this paradigm achieves near-perfect results. Yet state-of-the-art systems \cite{foscarin2024beat} score as low as $F=0.63$ on the SMC dataset \cite{holzapfel2012selective}.

Low scores in SMC are unsurprising in isolation, as the dataset was deliberately constructed from musically challenging excerpts. However, the wide gap between near-ceiling performance on standard benchmarks and persistent failure on SMC suggests that current architectures have systematic blind spots. Understanding exactly where and why models fail, rather than reporting a single aggregate score for the entire dataset, is a necessary step toward addressing these limitations.

In this paper, we present the first fine-grained diagnostic analysis of beat tracking performance on the SMC dataset, leveraging the per-track difficulty tags that have gone unused since the dataset's introduction. We evaluate three state-of-the-art systems: Beat This, \texttt{DBNBeatTracker} from madmom, and Beat Transformer. We then organize the dataset's 23 difficulty descriptors into four categories/axes of musical difficulty: weak beat cues, tempo instability, metrical ambiguity, and structural difficulty. Through this lens, we identify three distinct failure modes: octave errors (correct phase but wrong metrical level), continuity errors (locally correct but globally unstable tracking), and complete tracking failure ($F<0.3$ and $\textrm{AMLt}<0.3$).

The analysis of raw activation functions reveals that the dominant failure mode is not weak or absent activations, but that the primary bottleneck is the neural network's activation function itself: models trained on metronomically stable music produce confident but incorrect activations on expressively timed music. Setting the \texttt{min\_bpm} and \texttt{max\_bpm} values of the DBN to the ground-truth tempo $[-20\%, +20\%]$ for each song improves metrical coherence (CMLt) but does not improve beat placement (F-measure), showing that these are independent problems.

We also find that lowering the DBN's minimum tempo from 55 to 30\,BPM allows it to infer the correct tempo for the 45 tracks (21\%) that are otherwise forced into double-tempo predictions. More broadly, our analysis reveals that beat tracking failure on SMC is driven primarily by the activation function, not the DBN; that tempo instability is the dominant cause for beat prediction errors; and that current systems cannot estimate tempo accurately at inference time, leaving a substantial CMLt gap that correct tempo information could close. Furthermore, we show that treating the DBN's tempo-continuity hyperparameter as adaptive rather than fixed yields $F = 0.642$ when optimized per track, surpassing Beat This's raw peak-picking score of $F = 0.627$. No single fixed setting can serve both populations, as tracks with clean activations and tracks with degraded activations require opposing configurations.

\section{Background}

\subsection{Dynamic Bayesian Networks}

Most beat tracking models do not directly produce a final sequence of beats and downbeats but rather produce a temporal activation function, in which the values at each time index refer to the probability that a beat is present at that location. The activation function is then provided to a Dynamic Bayesian Network (DBN), which can then be used to determine the most likely beat and downbeat sequence. The original bar pointer model \cite{whiteley2006bayesian} was most notably expanded by \cite{krebs2015efficient} and \cite{bock2016joint}. Although certain approaches \cite{steinmetz2021wavebeat, chen2022toward, foscarin2024beat, ahn2025beat} have attempted to omit the DBN, each with varying results, a consistent pattern shows that global tempo consistency constraints provided by the DBN generally lead to better results.

\subsection{Evaluation metrics}
The primary evaluation metric is the F-measure, computed within a $\pm$70\,ms tolerance window around each ground truth beat~\cite{davies2009evaluation, dixon2001automatic, raffel2014mir_eval}. We also report continuity-based metrics. CMLt and AMLt are defined as the ratio of the longest continuously correctly-tracked segment to the total number of annotations, at the correct metrical level and at any allowed metrical level, respectively. Their continuous-segment variants (CMLc, AMLc) report this longest segment as a fraction of the total track length. High AMLt with low CMLt indicates tracking at double or half the annotated tempo; high CMLt with low CMLc indicates correct local tracking that is frequently interrupted during complex passages.

\section{Revisiting the SMC dataset}\label{sec:smc}

The SMC dataset \cite{holzapfel2012selective} is a beat tracking dataset containing 217 manually annotated 40-second Western music excerpts specifically compiled to evaluate beat tracking algorithms on rhythmically complex audio. The audio is formatted as mono WAV files sampled at 44.1 kHz. The excerpts were selected using a Query-by-Committee approach \cite{seung1992query}: multiple beat tracking algorithms were run on 678 excerpts, and the 270 excerpts with the lowest inter-algorithm agreement (Mean Mutual Agreement $\le$ 1 bit) were retained as the `hard' subset. A control group of 19 (`easy') excerpts with the highest mutual agreement scores were also included in the dataset in order to validate that low mutual agreement actually correlated with beat tracking difficulty. Annotations were produced through a rigorous multi-stage process involving spontaneous tapping, manual labeling, and expert review; see \cite{holzapfel2012selective} for details.

The final SMC dataset consists not only of the audio and beat annotations, but also a text file for each excerpt containing free-text difficulty descriptors that explain the cause of difficulty in beat tracking (e.g., expressive timing, missing bass, or ternary meter). It also mentions which annotator labeled the beats and a confidence level for the beat annotation itself. The SMC dataset does not have downbeat labels.

\section{Experimental setup}

To isolate the causes of beat tracking failures on the SMC dataset, we evaluate using three representative architectures: Beat This \cite{foscarin2024beat}, a Transformer-based system that represents the current state-of-the-art; Beat Transformer \cite{zhao2022beat}; and madmom's \texttt{TCNBeatProcessor} \cite{bock2016madmom, matthewdavies2019temporal, bock2020deconstruct}. We employ a rigorous 8-fold cross-validation setup for the Transformer models to ensure every track is evaluated by a checkpoint that held it out during training.

A primary goal of this study is to separate the contribution of the activation function from that of the DBN. This separation is necessary because the DBN's rigid tempo prior can shift beats away from where the activation function places them, sometimes worsening the result. To observe this, we evaluate Beat This using both its native raw peak-picking and the standard \textit{madmom} DBN. Unless otherwise stated, all DBN-based experiments utilize the library-default $[55, 215]$ BPM range. Evaluation is conducted using the \texttt{mir\_eval.beat.evaluate} function. We replicate Beat Transformer's published $F=0.596$ by converting its frame indices to timestamps at 43.07 FPS and by evaluating the full track rather than applying \texttt{mir\_eval}'s default 5-second trim.

\section{Results \& analysis}

\begin{table*}[t]
\centering
\small
\newcolumntype{L}[1]{>{\raggedright\arraybackslash}p{#1}}
\newcolumntype{C}[1]{>{\centering\arraybackslash}p{#1}}
\setlength{\tabcolsep}{4pt}
\begin{tabular}{@{}L{2.5cm} L{4.5cm} r C{0.8cm} C{0.5cm} r C{0.7cm} C{0.7cm} C{1.2cm} C{1.2cm}@{}}
\toprule
 & & & \multicolumn{3}{c}{Act@GT} & \multicolumn{2}{c}{Peak vs. DBN} & \multicolumn{2}{c}{GT-tempo $\pm 20\%$} \\
\cmidrule(lr){4-6} \cmidrule(lr){7-8} \cmidrule(lr){9-10}
Axis & Top tags & {$n$} & {On-axis} & {Off-axis} & {$\rho$} & {$\Delta$F} & {\% hurt} & {$\Delta$CMLt (on-axis)} & {$\Delta$CMLt (off-axis)} \\
\midrule
Weak beat cues     & missing bass (72),\newline lack of transients (70), +5 others     & 140 & .678 & .687 & $-$.040 & $-$.045 & 54\% & $+$.169 & $+$.218 \\
Tempo instability  & expressive\ timing (124),\newline slow tempo (71), +3 others           & 165 & .652 & .774 & $-$.305* & $-$.063 & 65\% & $+$.171 & $+$.233 \\
Metrical ambiguity & ternary meter (70),\newline strong syncopation\ (24), +3 others       & 103 & .655 & .705 & $-$.127 & $-$.057 & 61\% & $+$.194 & $+$.179 \\
Structural/ context & rich ornamentation\ (25),\newline low familiarity (23), +2 others       &  62 & .667 & .687 & $-$.049 & $-$.061 & 65\% & $+$.149 & $+$.201 \\
\bottomrule
\multicolumn{10}{l}{\footnotesize Full tag listings are provided as supplementary material.}
\end{tabular}
\caption{Four difficulty axes derived from SMC's normalized tags, analyzed using Beat This under 8-fold cross-validation. Act@GT: mean Beat This activation value at ground truth beat positions; $\rho$: Spearman correlation between Act@GT and F-measure (* $p < 0.001$, all others $p > 0.05$). Peak-picking vs.\ DBN columns show the effect of routing Beat This activations through madmom's DBN: $\Delta$F is the mean F-measure change (DBN minus peak-picking) and \% hurt is the fraction of tracks where the DBN reduces F by $>$0.01. $\Delta$CMLt columns show the CMLt gain from constraining the DBN to $\pm$20\% of the ground truth tempo, reported separately for tracks assigned to each axis (on-axis) and the remaining tracks (off-axis). Full tag listings are provided as supplementary material.}

\label{tab:axes}
\end{table*}

\subsection{SMC dataset analysis} \label{sec:smc_analysis}

After normalizing spelling variants, singular/plural forms, and parenthesized duplicates in the per-track \texttt{.tag} files, the dataset contains 23 unique difficulty descriptors. We grouped these tags into four axes based on the type of challenge each presents: \textit{weak beat cues}, with absent or faint acoustic beat markers; \textit{tempo instability}, or non-constant beat periods; \textit{metrical ambiguity}, or competing valid beat interpretations; and \textit{structural difficulty} (see Table~\ref{tab:axes} for details). When running Beat This using the four axes, we realized that SMC's difficulty is fundamentally combinatorial. Hard tracks have on average 3.85 difficulty tags and activate 2.33 of the four difficulty axes, compared to 1.21 tags and 0.42 axes for easy tracks. Among hard tracks, 84\% activate two or more axes simultaneously. Mean F-measure generally decreases with the number of active axes: 0.782 (0 axes, $n$=16), 0.645 (1, $n$=33), 0.634 (2, $n$=82), 0.583 (3, $n$=71), and 0.593 (4, $n$=15). The trend is sharpest between zero and one active axis; once multiple axes are present, the effect of adding more largely plateaus.

\begin{figure}[t]
\centering
\includegraphics[width=\columnwidth]{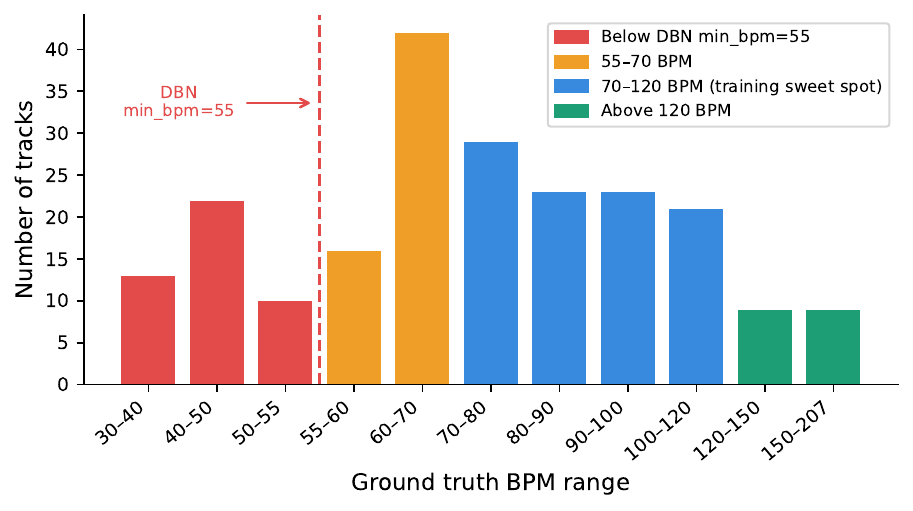}
\caption{Ground truth tempo distribution of SMC. The dashed line marks madmom's default DBN minimum of 55~BPM; the 45 tracks (21\%) to its left cannot be represented at the correct tempo under default settings.}
\label{fig:bpm-dist}
\end{figure}

SMC is also heavily skewed toward slow tempos (Figure~\ref{fig:bpm-dist}): median 70.9~BPM, IQR (middle 50\%) 57--93~BPM, with 45 tracks (21\%) below 55~BPM and 61 (28\%) below 60~BPM. This contrasts sharply with standard evaluation benchmarks; Hainsworth \cite{hainsworth2004particle} and Ballroom \cite{gouyon2006experimental, krebs2013rhythmic} have average tempos of 114~BPM and 130~BPM, respectively. Beat This achieves F\,=\,0.627, CMLt\,=\,0.514, AMLt\,=\,0.610 on SMC under 8-fold cross-validation \cite{foscarin2024beat}. Easy tracks score F\,=\,0.819 versus F\,=\,0.609 for hard tracks.

\subsection{Baseline performance and failure taxonomy}

Performance degrades monotonically with annotator confidence ($F = 0.722$ at confidence 1 to $F = 0.421$ at confidence 4) and with descriptor count ($F = 0.732$ for 1 tag to $F = 0.413$ for 8).


We first classify tracks scoring F\,$\geq$\,0.8 as \textit{good}  (62~tracks, 29\%). Among the remaining 155 tracks, we identify \textit{octave errors} (AMLt\,$-$\,F\,$>$\,0.25; 17~tracks, 8\%), \textit{continuity errors} (CMLt\,$-$\,CMLc\,$>$\,0.2; 38~tracks, 18\%), and \textit{total failure} (F\,$<$\,0.3 \textit{and} AMLt\,$<$\,0.3; 11~tracks, 5\%), with the remaining 89~tracks showing moderate but uncharacteristic degradation.

\subsection{Activation diagnostic}\label{sec:activation}

Understanding the pieces of music that cause total failures or even mediocre average ground-truth activations does not directly inform us whether the issue is caused by weak activations, misplaced activations, or both. In order to understand the cause for total failures, we extract Beat This's raw beat activation functions for all 217 tracks and compute quantitative diagnostics: peak sharpness, beat periodicity strength, activation entropy, maximum activation within $\pm$2 frames of each ground truth beat position, and mean activation at false-positive positions.

\begin{figure}[t]
\centering
\includegraphics[width=\columnwidth]{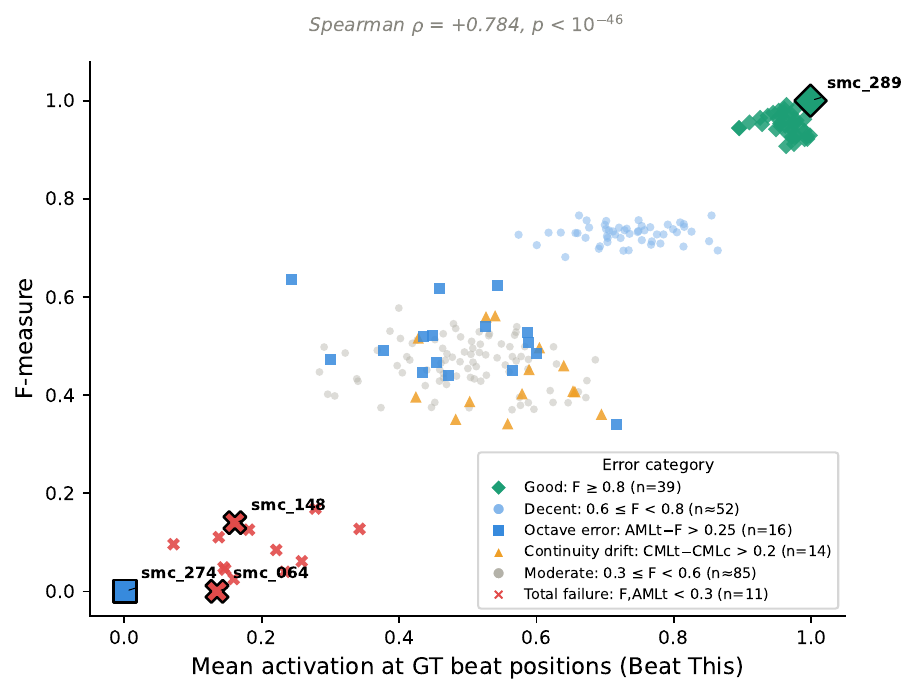}
\caption{Relationship between Beat This activation quality at GT beat positions and F-measure across all 217 SMC tracks. Marker shape encodes error category. The strong positive correlation ($\rho$\,=\,+0.784) confirms that activation at GT positions is the single best predictor of performance. Note that \texttt{smc\_274} (blue square near origin) has AMLt\,=\,0.988---Beat This fires at double tempo with offset phase, placing peaks between GT beats. Error categories in this figure use the intersection of Beat This and Beat Transformer performance (e.g., \emph{good} requires $F \geq 0.8$ on both systems), yielding smaller group sizes than the Beat-This-only taxonomy in Section~5.2.}
\label{fig:scatter}
\end{figure}

The total failure tracks produce strong activation peaks; they are simply at wrong positions. The maximum activation value across the track barely changes between tracks that the system handles well (0.998) and tracks where it fails completely (0.931). By contrast, the average activation at ground truth beat positions falls from 0.915 on good tracks to 0.284 on total failures, a 3.2× reduction (Figure~\ref{fig:scatter}). This metric is the single strongest correlate of F-measure across all 217 tracks (Spearman $\rho$\,=\,+0.784, $p$\,$<$\,$10^{-46}$). Of the 11 \textit{total failure} tracks, 10 activations exhibit \textit{wrong peaks} and just one is \textit{reasonable but misaligned}. These same 11 tracks also prove difficult for Beat Transformer.

Listening analysis of these tracks reveals that the wrong peaks correspond not to random noise but rather to genuine acoustic events: piano keystroke transients (\texttt{smc\_254}), guitar attack onsets (\texttt{smc\_158}, \texttt{smc\_203}), vocal entry points (\texttt{smc\_148}, \texttt{smc\_202}), and amplitude envelope changes (\texttt{smc\_064}). The model is not failing to detect \emph{something}; it is confidently detecting non-beat events that share the transient energy profile of percussive beats in its training data.

To test whether the activation function is the limiting factor, we ran two experiments on Beat This's activations. First, we checked whether occasional spurious or missing beats could be patched by flagging individual inter-beat intervals that deviated from the local median and removing or interpolating them. Second, we swept the peak-picking threshold from 0.05 to 0.98 to test whether the default threshold of 0.5 is suboptimal for SMC. None of these improved over the baseline. However, selecting the best threshold independently for each track yields $F = 0.673$, a +0.046 improvement that establishes an upper bound on what any DBN can achieve with these activations. The gap between this ceiling and the $F = 0.924$ achieved with GT activations (Section~\ref{sec:bottleneck}) confirms that the primary bottleneck is the activation function itself.

\subsubsection{Activation quality by difficulty axis}\label{sec:act-axis}

If the activations themselves are the bottleneck, a natural question is which musical properties cause the model to produce wrong peaks. The difficulty tags let us test this directly. To connect the activation diagnostic to the dataset metadata, we grouped tracks by the four difficulty axes and compared average activation at ground truth positions across groups (Table~\ref{tab:axes}).

\begin{table}[t]
\centering
\small
\setlength{\tabcolsep}{3.5pt}
\newcolumntype{C}[1]{>{\centering\arraybackslash}p{#1}}
\begin{tabular}{@{}lr C{0.7cm} C{0.9cm} C{0.9cm} C{0.9cm} C{0.7cm}@{}}
\toprule
 & & & \multicolumn{3}{c}{F-measure} & \\
\cmidrule(lr){4-6}
Dataset & $N$ & {IBI\newline CV} & {Real\newline +peak} & {Real\newline +DBN} & {GT\newline +DBN} & {Gap} \\
\midrule
Ballroom   & 685 & .021 & .986 & .965 & .922 & $-$.043 \\
Beatles    & 179 & .023 & .979 & .956 & .975 & $+$.019 \\
GTZAN      & 993 & .017 & .891 & .880 & .965 & $+$.085 \\
Hainsworth & 222 & .036 & .907 & .901 & .986 & $+$.085 \\
\textbf{SMC} & \textbf{217} & \textbf{.091} & \textbf{.627} & \textbf{.585} & \textbf{.924} & \textbf{$+$.339} \\
\bottomrule
\end{tabular}
\caption{Cross-dataset comparison of tempo variability and activation bottleneck. IBI CV: median coefficient of variation of ground truth inter-beat intervals (higher = more tempo variability). Gap: GT+DBN minus Real+DBN F-measure (all DBN results use min\_bpm=30). SMC has 2.5--5.3$\times$ higher tempo variability and a 4$\times$ larger activation bottleneck gap than any other dataset.}
\label{tab:bottleneck}
\end{table}

Tempo instability is the only axis with a statistically significant correlation to activation quality ($\rho$\,=\,$-$0.305, $p$\,$<$\,0.001). It also shows the largest F-measure gap ($-$0.113) and the sharpest degradation in peak sharpness (0.649 for tracks in the axis vs.\ 0.835 for the remaining) and periodicity (0.510 vs.\ 0.788). This means that tempo instability, which is the most common tag in SMC, causes failures at the activation level, not just the DBN level. As can be seen in Table~\ref{tab:bottleneck}, SMC's ground truth inter-beat intervals are $2.5-5.3\times$ more variable than those of standard training datasets, suggesting that models trained predominantly on metronomically stable music produce activations that degrade when confronted with rubato and tempo variation. Weak beat cues, despite being the most prevalent axis (140/217 tracks), shows essentially no correlation with activation quality ($\rho$\,=\,$-$0.040, $p$\,=\,0.55).

\subsubsection{Activation bottleneck quantification}
\label{sec:bottleneck}

To directly measure whether the bottleneck lies in the activation function or the DBN, we used the ground truth observation model. We created synthetic Gaussian peaks ($\sigma = 2\textrm{ frames}$, ~40 ms) centered at ground truth beat positions as the observation model and processed them with the standard DBN. With $\texttt{min\_bpm} = 30$, this achieved $F = 0.924$ across all 217 tracks, compared to 0.585 with real activations: a +0.339 improvement, with only 2 tracks remaining below F = 0.5. To test whether this gap is SMC-specific, we repeated the experiment on four standard datasets (Table~\ref{tab:bottleneck}): Ballroom \cite{gouyon2006experimental, krebs2013rhythmic}, Hainsworth \cite{hainsworth2004particle}, Beatles \cite{davies2009evaluation}, and GTZAN \cite{marchand2015swing}. The gap scales with dataset difficulty: +0.019 on Beatles, +0.085 on GTZAN and Hainsworth, and -0.043 on Ballroom, where the model already achieves $F = 0.986$ with simple peak-picking. On SMC, the gap is $4\times$ larger than any other dataset, confirming that the activation bottleneck is specific to SMC's combination of expressive timing, slow tempos, and weak beat cues.

\subsection{The role of tempo}\label{sec:tempo}

The aforementioned F-measure score ceiling of 0.924 confirms that the DBN is not the main bottleneck, suggesting that much of the remaining gap from current performance (0.585) reflects the DBN working with degraded input. A key question is whether providing the DBN with better tempo information could recover some of this gap.

\subsubsection{DBN: the tempo-continuity tradeoff}

Routing SMC beat activations from Beat This through madmom's DBN hurts F-measure by $-$0.051 (0.627\,$\to$\,0.576) while improving AMLt by +0.046 (0.610 $\to$ 0.656). Of 217 tracks, 127 are worsened, 62 improved, and 28 unchanged. Grouping by difficulty axis (Table~\ref{tab:axes}) reveals that the DBN hurts tempo instability tracks most severely ($-$0.063~F, 65\% worsened) while being nearly neutral on tracks \textit{without} tempo instability ($-$0.012~F, 29\% worsened). The DBN's rigid tempo-continuity prior is precisely wrong for expressive music but performs adequately when its assumption of temporal regularity holds.

\subsubsection{The effect of widening the DBN tempo range}\label{sec:wide_dbn}

madmom's DBN defaults to $\texttt{min\_bpm}=55$, but 45 SMC tracks ($21\%$) have ground truth tempo below that threshold (Figure~\ref{fig:bpm-dist}), forcing double-tempo predictions. Lowering $\texttt{min\_bpm}$ to 30 improves the 45 slow tracks (Beat Transformer $F: 0.381 \to 0.458$) with no effect on the remainder (Table~\ref{tab:bottleneck}).

\subsubsection{Tempo estimation accuracy} \label{sec:tempo_est}

Beat Transformer's tempo classification head achieves 59\% accuracy on SMC---2.6$\times$ better than madmom's standalone TCN accuracy rate of 22\%. Accuracy varies sharply by BPM range: 85\% at 70--90~BPM, 31\% below 55, 11\% above 120, directly mirroring training data coverage. Of the 41\% errors: 9.7\% are double tempo, 10.1\% half tempo, and 21.7\% other.


\subsubsection{Tempo-constrained DBN}\label{sec:tempo-constraint}

\begin{figure}[t]
\centering
\includegraphics[width=\columnwidth]{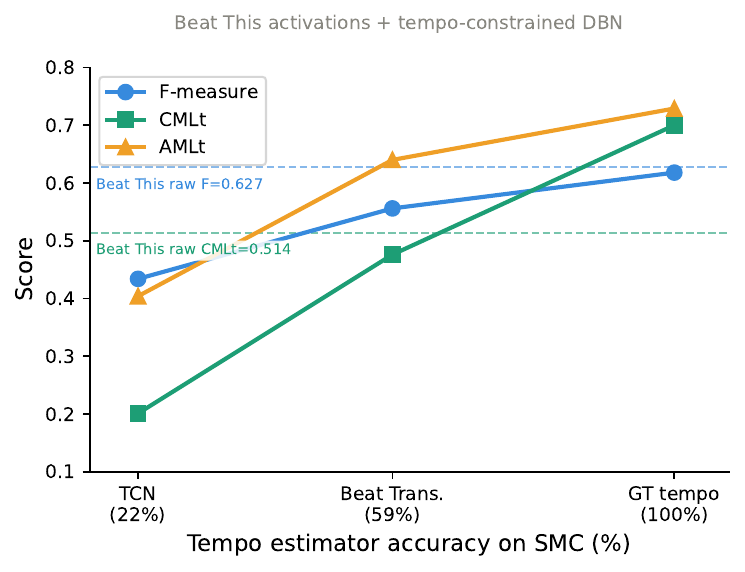}
\caption{Effect of tempo estimation accuracy on beat tracking performance. CMLt (green) rises sharply from 0.201 to 0.700 as tempo accuracy improves, but F-measure (blue) remains below Beat This's unconstrained baseline (dashed). This divergence is the central evidence that beat placement and metrical coherence are two independent problems on SMC.}
\label{fig:three-point}
\end{figure}

In the standard DBN, tempo is a latent variable inferred jointly with beat positions from the activation function, constrained to change only between bars \cite{krebs2015efficient, krebs2016downbeat}. To test whether providing better tempo information could help, we constrained the DBN's BPM range to a $\pm 20\%$ window around tempo estimates of increasing accuracy (Figure~\ref{fig:three-point}). CMLt rises sharply from 0.201 to 0.700 as tempo accuracy improves, while F-measure remains below Beat This's unconstrained baseline of 0.627. This divergence is the central evidence for two independent problems: beat \textit{placement} (measured by F) and metrical \textit{coherence} (measured by CMLt).

\subsubsection{GT-tempo upper bound: what perfect tempo can and cannot fix}\label{sec:gt-tempo}

Constraining the DBN to $\pm20\%$ of the ground truth tempo improves CMLt from 0.514 to 0.700 ($+$0.186) and AMLt from 0.610 to 0.729, but F-measure \textit{decreases} from 0.627 to 0.618. Raw peak-picking places individual beats more precisely than the DBN's rigid phase model, even with correct tempo.

Grouping by error category reveals what perfect tempo can and cannot fix. CMLt improves across every category, as perfect tempo helps coherence universally. But F-measure does not improve on the total failure tracks ($-$0.020) or continuity error tracks ($-$0.015). The most extreme case is \texttt{smc\_064}: $F=0.000$ with GT tempo but $\textrm{CMLt}=0.880$, in which the model places coherently wrong beats at the correct rate. The per-track optimal threshold experiment (Section \ref{sec:activation}) establishes an upper bound of $F=0.673$ for any DBN operating on the real activations, while GT activations achieve $F=0.924$. Of the 0.297 gap, 85\% ($F=0.251$) is attributable to wrong activation peaks, and 15\% ($F=0.046$) to the limits of the DBN.



Grouping by difficulty axis (Table~\ref{tab:axes}) shows that metrical ambiguity tracks benefit most from perfect tempo (+0.194 $\Delta$CMLt), making this the clearest ``tempo ceiling'' axis---tracks where activations exist at plausible beat positions but at the wrong metrical level, fixable with correct tempo. Tempo instability tracks show \textit{smaller} gains than for tracks not on the axis (+0.171 vs.\ +0.233), because their degraded activations (Section~\ref{sec:act-axis}) mean even perfect tempo cannot fully rescue them.

\begin{table}
  \centering
  \begin{tabular}{lccc}
    \hline
    Configuration & F & CMLt & AMLt \\
    \hline
    Peak-picking & .627 & .514 & .610 \\
    DBN $\lambda$=100 & .576 & .474 & .656 \\
    DBN optimal $\lambda$ & .642 & .558 & .637 \\
    GT-tempo + optimal $\lambda$ & \textbf{.667} & \textbf{.735} & \textbf{.755} \\
    \hline
    GT activations + DBN & .924 & .921 & .925 \\
    \hline
  \end{tabular}
  \caption{Beat This configurations on SMC. Per-track optimal $\lambda$ surpasses raw peak-picking (.642 vs .627); adding GT-tempo yields .667 F and .735 CMLt, confirming the two gains compound. Bottom row: upper bound with perfect activations.}
  \label{tab:systems}
\end{table}

\subsubsection{DBN parameter sensitivity}\label{sec:lambda-sweep}

The DBN's \texttt{transition\_lambda} parameter controls the tradeoff between placement precision and metrical coherence: high values enforce rigid tempo continuity, low values allow the DBN to follow the activation more freely. Thus, we swept \texttt{transition\_lambda} across 13 values from 1 to 500 for each of the 217 tracks, keeping all other parameters fixed (\texttt{min\_bpm}=30, \texttt{max\_bpm}=215). For each track, we retained the $\lambda$ maximizing F-measure. Using this approach, we achieved F\,=\,0.642 and CMLt\,=\,0.558, a +0.050 F improvement over the library default of $\lambda$\,=\,100 (F\,=\,0.592)---larger than the gap between any two systems we evaluated. Notably, this is the first DBN configuration in our experiments that surpasses Beat This's raw peak-picking on F-measure (0.642 vs.\ 0.627) while also improving CMLt (0.558 vs.\ 0.514).

The distribution of the optimal $\lambda$ values is strongly bimodal. One-third of these tracks (73/217) prefer $\lambda$\,=\,1, meaning minimal tempo smoothing; the DBN's tempo prior actively hurts these tracks. Only 5\% of tracks prefer the default $\lambda$\,=\,100. The globally optimal fixed $\lambda$ is 5, not 100, yielding F\,=\,0.603 (+0.011 over default).

Tracks classified as \textit{good} (F\,$\geq$\,0.8 without DBN) overwhelmingly prefer $\lambda$\,=\,1 (median optimal $\lambda$\,=\,2): their activations are already clean, and tempo smoothing only introduces errors. In contrast, \textit{total failure} and \textit{octave error} tracks prefer higher values (median 30 and 20, respectively), consistent with noisy activations benefiting from stronger priors. This pattern reveals a fundamental tension in the fixed-parameter DBN: the setting that helps tracks with poor activations is precisely the setting that damages tracks with good activations. No single $\lambda$ can serve both populations, and the +0.050 F gap quantifies the cost of this compromise. This experiment ultimately shows the DBN tempo transition model is too rigid.

\section{Discussion}

\subsection{Two performance ceilings}

Our experiments converge on two independent performance ceilings on SMC. The activation ceiling ($\sim$F\,=\,0.67) is the maximum F-measure achievable across all system and DBN combinations; it exists because approximately 100 tracks produce confident activation peaks at wrong positions that no DBN can override. 
The tempo ceiling ($\sim$CMLt\,=\,0.70 with fixed $\lambda$, rising to 0.735 with per-track optimal $\lambda$; Table~\ref{tab:systems}) is the maximum metrical coherence achievable with perfect tempo through the current DBN, requiring accurate tempo estimation and a more flexible integration mechanism than hard BPM constraints. Metrical ambiguity maps neatly to the tempo ceiling, while tempo instability hits both ceilings simultaneously, degrading the activation function and the DBN alike. As shown in Table \ref{tab:systems}, combining per-track optimal $\lambda$ with GT-tempo reaches $F=0.667$ and $\textrm{CMLt}=0.735$, approaching both ceilings simultaneously, but only when the best $\lambda$ and the correct tempo are known for each track. Some portion of the activation ceiling may reflect genuine metrical ambiguity rather than model failure, particularly on tracks with annotator confidence 3--4.

\subsection{Training data as the likely cause of the activation ceiling}

The activation ceiling likely reflects training data mismatch. Beat This trains on 16 datasets dominated by percussion-driven, steady-tempo music, with zero acapella material and a tempo distribution clustered in 80--140 BPM, far from SMC's median of 70.9 BPM (Section \ref{sec:smc_analysis}). Two training-time interventions could help. Data augmentation via source separation, such as removing drums from existing training datasets rather than just providing them as separate input streams \cite{zhao2022beat}, could force the model to learn non-percussive beat cues without requiring new annotations. Loss function redesign may also help: the activation diagnostic shows the model's problem is not missing beats but confidently predicting wrong ones, suggesting that a loss penalizing high-confidence predictions far from ground truth could reshape the activation landscape.

\subsection{Tempo at inference time}

Our experiments identify two independent levers for improving beat tracking on SMC. First, as shown in Section \ref{sec:gt-tempo}, providing the DBN with accurate tempo information---rather than discarding the tempo prediction as a training regularizer---could recover much of the +0.186 CMLt gap demonstrated by injecting the ground-truth tempo. Second, as shown in Section \ref{sec:lambda-sweep}, allowing $\lambda$ to vary per track recovers +0.050 F over the library default. Table \ref{tab:systems} confirms that these two gains compound: GT-tempo with per-track optimal $\lambda$ achieves $F=0.667$ and $\textrm{CMLt}=0.735$, the best result in any configuration we tested. Since one lever improves metrical coherence while the other improves beat placement, a learned DBN that takes the  predicted tempo as an inference-time input and adapts its tempo-continuity prior to the musical context could capture both gains simultaneously. Notably, prior systems \cite{bock2020deconstruct, zhao2022beat} jointly train a tempo head but discard it at inference, using it only as a training regularizer; our GT-tempo experiment shows this leaves a +0.186 CMLt gain on the table.

\section{Conclusion}

We presented the first per-track diagnostic analysis of beat tracking on the SMC dataset \cite{holzapfel2012selective}. Our analysis identifies three distinct failure modes and reveals that the dominant cause of low performance is confident-but-wrong activation peaks. The per-track optimal threshold experiment places an upper bound of $F = 0.673$ on any decoder operating on the real activations, while GT activations achieve $F = 0.924$, confirming that the majority of the performance gap originates in the activation function. Tempo instability is the only difficulty axis that degrades activation quality itself ($\rho = -0.305$, $p < 0.001$), while metrical ambiguity is the clearest target for tempo-aware inference ($+0.194$~$\Delta$CMLt with perfect tempo). Combining per-track optimal $\lambda$ with GT-tempo reaches $F = 0.667$ and CMLt $= 0.735$, confirming that these two gains are independent and additive. These results point toward two complementary improvements: diversifying training data to address the activation ceiling, and replacing the fixed-parameter DBN with a learned version that adapts both its tempo estimate and tempo-smoothness prior probability to each track's musical context.

\section{AI Usage Statement}

We declare the following use of AI tools in the preparation of this manuscript. Claude (Anthropic) was used as a writing assistant for drafting and revising manuscript prose, structuring the presentation of results, and reviewing the manuscript for numerical inconsistencies. Claude Code (Anthropic) was used to verify reported figures against experimental source code and to trace discrepancies between code outputs and hardcoded values in draft text. All experimental design, implementation, data analysis, listening analysis, and analytical conclusions are the authors' own work. All AI-generated text was reviewed, verified, and edited by the authors, who take full responsibility for the content of this paper.

\bibliography{ISMIRtemplate}

%
%
%
%

\end{document}